\documentclass[prb,twocolumn,reprint,showpacs,superscriptaddress,floatfix]{revtex4-1}
\usepackage{graphicx}
\usepackage{epsfig}
\usepackage{epstopdf}
\begin{document}

\title{Magnetic interactions in the Martensitic phase of Mn rich Ni-Mn-In shape memory alloys}
\author{D. N. Lobo}
\address{Department of Physics, Goa University, Taleigao Plateau, Goa 403 206 India}
\author{Sandhya Dwivedi}
\address{Tata Institute of Fundamental Research, Dr. Homi Bhabha Road, Colaba, Mumbai 400 005 India}
\author{C. A. daSilva}
\author{N. O. Moreno}
\address{Departamento de Fisica, Universidade Federal de Sergipe, S$\tilde{\rm a}$o Crist$\acute{\rm o}$v$\tilde{\rm a}$o, SE, 49100-000,
Brazil}
\author{K. R. Priolkar} \email[corresponding author: ]{krp@unigoa.ac.in}
\address{Department of Physics, Goa University, Taleigao Plateau, Goa 403 206 India}
\author{A. K. Nigam}
\address{Tata Institute of Fundamental Research, Dr. Homi Bhabha Road, Colaba, Mumbai 400 005 India}

\date{\today}

\begin{abstract}
The magnetic properties of Mn$_{2}$Ni$_{(1+x)}$In$_{(1-x)}$  ($x$ = 0.5, 0.6, 0.7) and
Mn$_{(2-y)}$Ni$_{(1.6+y)}$In$_{0.4}$ ($y$ = -0.08, -0.04, 0.04, 0.08) shape memory alloys have been studied. Magnetic interactions in the martensitic phase of these alloys are found to be quite similar to those in Ni$_2$Mn$_{(1+x)}$In$_{(1-x)}$ type alloys. Doping of Ni for In not only induces martensitic instability in Mn$_2$NiIn type alloys but also affects magnetic properties due to a site occupancy disorder. Excess Ni preferentially occupies X sites forcing Mn to the Z sites of X$_2$YZ Heusler composition resulting in a transition from ferromagnetic ground state to a state dominated by ferromagnetic Mn(Y) - Mn(Y) and antiferromagnetic Mn(Y)-Mn(Z) interactions.  These changes in magnetic ground state manifest themselves in observation of exchange bias effect even in zero field cooled condition and virgin magnetization curve lying outside the hysteresis loop.
\end{abstract}


\maketitle

\section{Introduction}
Interest in ferromagnetic shape memory alloys (FSMAs) due to their technological potential have generated much attention  in recent years \cite{Ullakko,Planes,Vasilev}. These alloys have the generic formula X$_2$YZ where the X and Y atoms are 3d elements while Z is a group IIIA-VA element. They undergo a diffusionless transformation from high temperature austenitic cubic structure to low temperature (martensitic) tetragonal or orthorhombic structure \cite{Vasilev}. One such system  extensively studied is  Ni$_2$MnGa which undergoes  martensitic transformation at T$_M$ $\sim$ 220K. Martensitic transformation can also be induced in Z = In, Sn or Sb but in off-stoichiometric compositions, Ni$_{2}$Mn$_{(1+x)}$Z$_{(1-x)}$ \cite{Krenke}. Apart from the martensitic transformation these alloys exhibit other interesting properties like large magnetocaloric effects, magnetic superelasticity \cite{Krenke,APlanes,Shamberger,Bhobe} and magnetic field induced giant strains \cite{Kainuma,Koyama}, exchange bias \cite{pat,pab} etc.

Alloys of the type Mn$_2$NiZ have higher Mn content compared to the traditional Ni$_2$MnZ and is considered beneficial in realizing better magnetic, magnetocaloric and magnetotransport properties \cite{koy,zho,li}. Mn$_2$NiGa is one such Mn rich alloy which undergoes martensitic transformation at T$_M$ $\sim$ 270 K in an ferrimagnetically ordered state (T$_C$ = 588K) \cite{liu}. Ferrimagnetic order has also been confirmed from band structure calculations and arise due to unequal magnetic moments of antiferromagnetically coupled Mn atoms occupying the X and Y sites of X$_2$YZ Heusler structure \cite{GDLiu,gdliu,Barman,barman,Souvik}. Although martensitic transformations have been theoretically predicted in other Mn$_2$NiZ (Z = In, Sn, Sb) alloys, experimental investigations have reported these alloys to have stable crystal structures \cite{Chakrabarti,Luo}. However, just as in case of Ni$_2$MnZ alloys wherein martensitic instability is caused by partial substitution of Z atoms by Mn, increasing of Ni content at the expense of Z atoms in Mn$_2$NiZ results in martensitic alloys \cite{LMa,Sanchez}. However, realization of such alloys with general composition Mn$_2$Ni$_{(1+x)}$Z$_{(1-x)}$ can lead to a structural disorder due to site preferences of transition metal ions.  In a L2$_1$ Heusler composition Ni atoms prefer X sites as compared to Z sites \cite{fel} and therefore doping excess Ni could result in newer magnetic interactions as it would force Mn to occupy the Z sites. These conditions can change the sign of RKKY interaction leading to magnetic frustration or a new type of magnetic order. Local structural disorder is shown to be primarily responsible for martensitic transformation and magnetic interactions in the martensitic state of Ni$_2$Mn$_{1+x}$In$_{1-x}$ type alloys \cite{Lobo,Priolkar,xmcd}. In case of Mn$_2$NiGa as well, site occupancy disorder has been shown to be an important factor in explaining magnetic properties of the martensitic state\cite{singh}. The antisite disorder is also shown to be responsible for the zero field cooled exchange bias in Mn$_2$PtGa \cite{nayak}.

In this paper we focus our attention on understanding the magnetic properties of
Mn$_{2}$Ni$_{(1+x)}$In$_{(1-x)}$ ($x$ = 0.5, 0.6, 0.7) and Mn$_{(2-y)}$Ni$_{(1.6+y)}$In$_{0.4}$ ($-0.08 \le y \le 0.08$) type alloys especially in their martensitic state. By comparing the magnetic properties of martensitic and structurally stable alloys having nearly similar compositions, we show that low temperature magnetic properties of the martensitic alloys are dominated by ferromagnetic and antiferromagnetic interactions, arising due to site occupancy disorder in these Mn rich alloys.

\section{Experimental}
The samples of above composition were prepared by arc melting the weighed constituents in argon atmosphere followed by encapsulating in a evacuated quartz tube and annealing at 750 C for 48 hours and subsequent quenching in ice cold water. The prepared alloys were cut in suitable sizes using a low speed diamond saw and part of the sample was powdered and re-annealed in the same procedure above. X-ray diffraction (XRD) patterns were recorded at room temperature in the angular range of $20^\circ \leq 2\theta \leq 100^\circ$. Resistivity was measured by standard four probe technique using a closed cycle helium refrigerator in  the temperature range from 10 K to 330K. Magnetization measurements were performed in the temperature interval 5 K - 400 K using a SQUID magnetometer. Here the samples were first cooled from room temperature to 5K in zero applied magnetic field and the data was recorded while warming (ZFC) followed by cooling (FCC) and subsequent warming (FCW) in the same applied field of 100 Oe. For high temperature magnetization vibrating sample magnetometer was employed in the temperature range of 300 K - 700 K. Isothermal magnetization M(H) at various temperatures were recorded by ramping the field in the range $\pm$8T in all five quadrants. All the M(H) loops were recorded by cooling the samples from room temperature to the desired temperature in zero applied field.

\section{Results}
\begin{figure}
\centering
\includegraphics[width=\columnwidth]{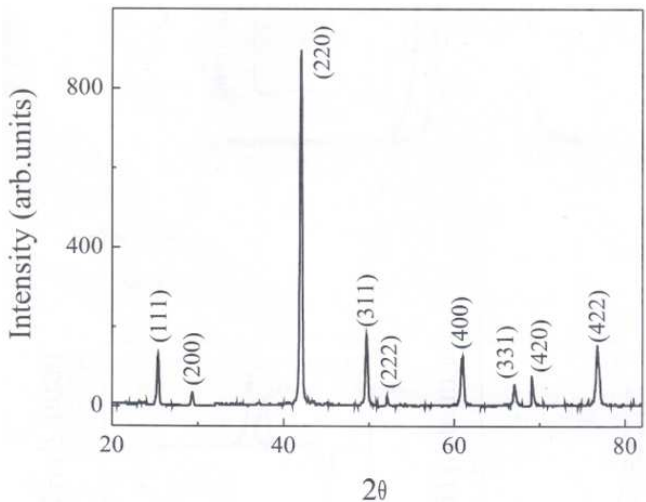}
\caption{\label{xrd.eps}XRD pattern for Mn$_{2}$Ni$_{1.6}$In$_{0.4}$ }
\end{figure}

%Mn$_2$Ni$_{1-x}$In$_{1-x}$ (x=0.5, 0.6, 0.7)
In  Fig. \ref{xrd.eps} a typical XRD pattern of  Mn$_{2}$Ni$_{1.6}$In$_{0.4}$ is shown. All the alloys except $x$ = 0.7 were found to be single phase with face centered cubic structure. The (111) and (200) peaks is an indication of the highly ordered Hg$_{2}$CuTi structure belonging to the  $Fm\bar3m$ space group.  XRD pattern of Mn$_{2}$Ni$_{1.7}$In$_{0.3}$ showed  mixed phase with reflections belonging to austenitic cubic and martensitic tetragonal phase indicating that the martensitic transformation temperature (T$_M$) of this alloy to be in the vicinity of room temperature. Indeed as shown later, the transformation temperature for this alloy is 350K.

\begin{figure}
\centering
\includegraphics[width=\columnwidth]{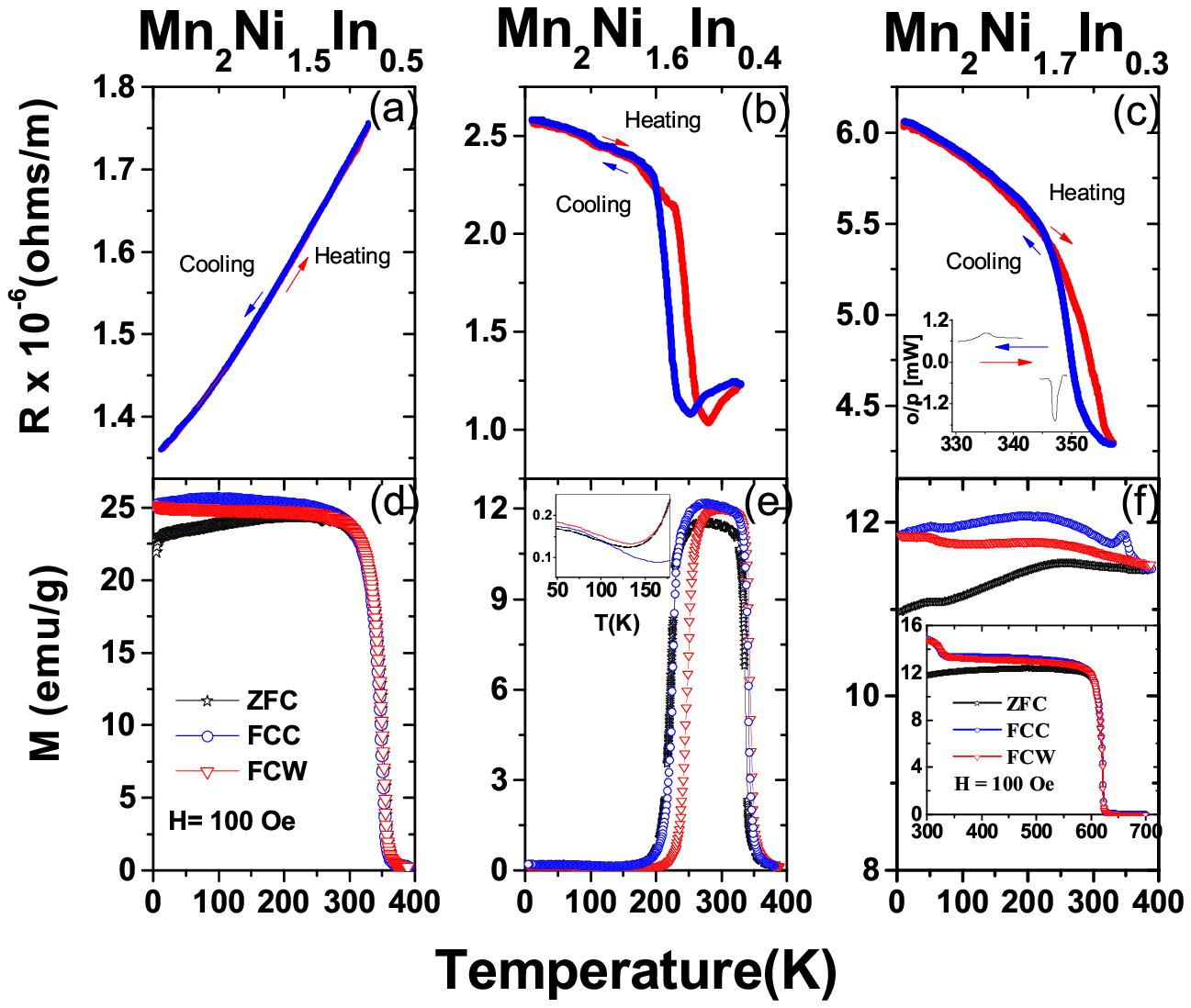}
\caption{\label{R-M2.eps} Resistivity plots for (a) Mn$_{2}$Ni$_{1.5}$In$_{0.5}$ (b) Mn$_{2}$Ni$_{1.6}$In$_{0.4}$ (c) Mn$_{2}$Ni$_{1.7}$In$_{0.3}$  and  Magnetization plots for (d) Mn$_{2}$Ni$_{1.5}$In$_{0.5}$ (e) Mn$_{2}$Ni$_{1.6}$In$_{0.4}$ (f) Mn$_{2}$Ni$_{1.7}$In$_{0.3}$  }
\end{figure}

In Fig. \ref{R-M2.eps}(a -f), plots of resistivity and magnetization (M(T)) as a function of temperature are presented for Mn$_2$Ni$_{1+x}$In$_{1-x}$ ($x$ = 0.5, 0.6, 0.7). It can be clearly seen from the nature of resistivity and magnetization curves that the two alloys $x$ = 0.6 and 0.7 undergo a first order magnetic transformation while $x$ = 0.5 has a stable ferro/ferrimagnetic metallic ground state. It may be noted that magnetic ordering in austenitic state has been termed ferrimagnetic based on band structure calculations on Mn$_2$NiIn wherein antiferromagnetic alignment of two Mn atoms with unequal moments has been predicted \cite{Chakrabarti,Luo}. The resistivity for alloys with $x$ = 0.6 and 0.7 shown in Figs. \ref{R-M2.eps}(b) and (c) show a sharp increase and a strong hysteresis between cooling and warming data at about 230K and 350K respectively. This is a signature of a transformation from high temperature austenitic phase to martensitic phase. The nature and transformation temperature has been further confirmed by differential scanning calorimetric measurements as depicted in inset of Fig. \ref{R-M2.eps}(c). Distortion of crystal structure in the region of martensitic transformation leads to changes in band structure that results in increased scattering of conduction electrons and therefore higher resistivity. Magnetization of these two alloys measured during FCC and FCW cycles also exhibits hysteresis indicating presence of a first order transformation as can be seen in Fig. \ref{R-M2.eps}(e) and (f). Another interesting aspect of magnetization is the near complete destruction of magnetic order upon martensitic transformation especially in Mn$_2$Ni$_{1.6}$In$_{0.4}$ (Fig. \ref{R-M2.eps}(e)). This effect is very similar to that observed in Mn rich Ni-Mn-In alloys \cite{KrenkeT}. The values of ferrimagnetic ordering temperature T$_C$ and T$_M$ estimated from the magnetization and resistivity data are tabulated in Table \ref{result}.

A closer examination of magnetization curves in Mn$_2$Ni$_{1.6}$In$_{0.4}$ reveals that below 125K, the magnetization starts increasing with decreasing temperature which is suggestive of another magnetic transition, T$_C^M$ typically referred to as magnetic ordering temperature of martensitic phase. Interestingly, below T$_C^M$, a splitting of the ZFC and the FC curves is observed which can be ascribed to the presence of competing antiferromagnetic interactions \cite{Xuan,JLSanchez,Chu}.

From the above results it seems that the martensitic transformation in Mn$_2$Ni$_{(1+x)}$In$_{(1-x)}$ alloys occurs only in alloys with  $x$ $\geq$ 0.6. This corresponds to an average valence electron per atom (e/a ratio) of 7.8. Interestingly, even in Ni$_{2}$Mn$_{(1+x)}$In$_{(1-x)}$ alloys , martensitic transformation  occurs in alloy composition with e/a ratio $\geq$ 7.8 \cite{KrenkeT}. In order to investigate the range of e/a ratio over which Mn-Ni-In alloys undergo martensitic transformation, the compositions were fine tuned by varying Mn to Ni ratio in Mn$_{2}$Ni$_{1.6}$In$_{0.4}$ resulting in alloys of the type Mn$_{(2+y)}$Ni$_{(1.6-y)}$In$_{0.4}$ (y = -0.08, -0.04, 0.04, 0.08).

\begin{figure}
\centering
\includegraphics[width=\columnwidth]{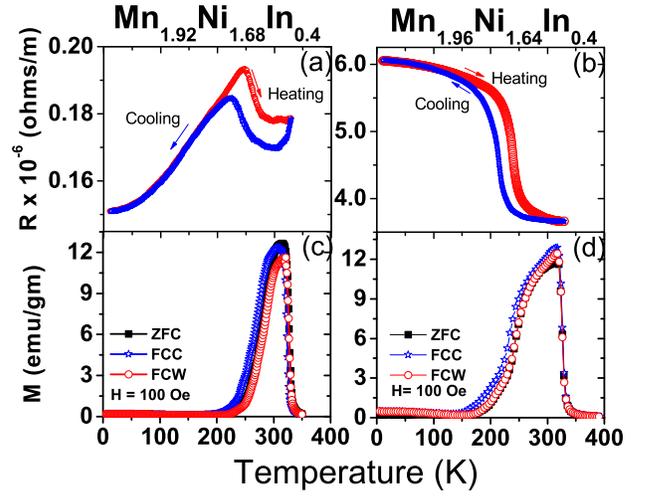}
\caption{\label{R-M3.eps} Resistivity plots for (a) Mn$_{1.92}$Ni$_{1.68}$In$_{0.4}$
(b)Mn$_{1.96}$Ni$_{1.64}$In$_{0.4}$ and Magnetization plots for (c) Mn$_{1.92}$Ni$_{1.68}$In$_{0.4}$
(d)Mn$_{1.96}$Ni$_{1.64}$In$_{0.4}$ }
\end{figure}

\begin{figure}
\centering
\includegraphics[width=\columnwidth]{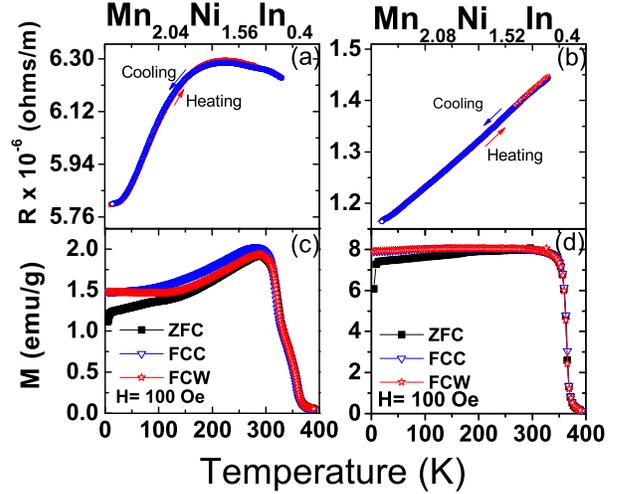}
\caption{\label{R-M4.eps} Resistivity plots for (a) Mn$_{2.04}$Ni$_{1.56}$In$_{0.4}$ and (b)
Mn$_{2.08}$Ni$_{1.52}$In$_{0.4}$ and Magnetization plots for (c) Mn$_{2.04}$Ni$_{1.56}$In$_{0.4}$ and (d)
Mn$_{2.08}$Ni$_{1.52}$In$_{0.4}$}
\end{figure}

Results of temperature dependence of resistivity recorded during warming and cooling cycles and magnetization recorded during ZFC, FCC and FCW cycles for Ni rich compositions Mn$_{1.92}$Ni$_{1.68}$In$_{0.4}$ and Mn$_{1.96}$Ni$_{1.64}$In$_{0.4}$ are presented in Fig. \ref{R-M3.eps}(a)-(b) and Fig. \ref{R-M3.eps}(c)-(d) respectively. Both the alloys are martensitic with T$_M$ 270K and 245K respectively. The behavior of magnetization is also very similar to that of Mn$_2$Ni$_{1.6}$In$_{0.4}$. It can also be noticed that with the increase in Ni concentration the martensitic transformation temperature increases. Whereas the Mn rich compositions, Mn$_{2.04}$Ni$_{1.56}$In$_{0.4}$ and Mn$_{2.08}$Ni$_{1.52}$In$_{0.4}$ exhibit ferrimagnetic metallic behavior (Fig. \ref{R-M4.eps}) similar to Mn$_2$Ni$_{1.5}$In$_{0.5}$. The magnetic and martensitic transformation temperatures for all the alloys are  listed in Table \ref{result}. It can be clearly seen that while magnetic ordering temperature decreases (except Mn$_2$Ni$_{1.7}$In$_{0.3}$), martensitic transformation temperature increases systematically with (e/a) ratio.

The nature of magnetization in Mn$_{1.92}$Ni$_{1.68}$In$_{0.4}$ (Fig\ref{R-M3.eps}(c)) and Mn$_{1.96}$Ni$_{1.64}$In$_{0.4}$ (Fig. \ref{R-M3.eps}(d)), is also very similar to that of Mn$_2$Ni$_{1.6}$In$_{0.4}$. In the martensitic phase, a splitting of the ZFC and FC curves is observed below T$_C^M$ $\sim$ 150K and 100K respectively, which again suggests the presence of competing magnetic interactions.

\begin{table}
\caption{\label{result}Values of e/a, T$_C$, T$_M$ and T$_C^M$ for Mn$_{2}$Ni$_{(1+x)}$In$_{(1-x)}$ and Mn$_{(2-y)}$Ni$_{(1.6+y)}$In$_{0.4}$ alloys }
\begin{tabular}{|c|c|c|c|c|}
\hline
Alloy & e/a ratio & T$_M$  & T$_C$ & T$_C^M$ \\
\hline
Mn$_2$Ni$_{1.7}$In$_{0.3}$ & 7.975 & 350K & 620K & -- \\
Mn$_{1.92}$Ni$_{1.68}$In$_{0.4}$ & 7.86 & 270K & 327K & 135K\\
Mn$_{1.96}$Ni$_{1.64}$In$_{0.4}$ & 7.83 & 245K & 326K & 125K \\
Mn$_2$Ni$_{1.6}$In$_{0.4}$ & 7.8 & 232K & 344K & 122K \\
Mn$_{2.04}$Ni$_{1.56}$In$_{0.4}$ & 7.77 & -- & 340K & --\\
Mn$_{2.08}$Ni$_{1.52}$In$_{0.4}$ & 7.74 & -- & 362K & --\\
Mn$_2$Ni$_{1.5}$In$_{0.5}$ & 7.625 & -- & 350K & --\\ \hline
\end{tabular}
\end{table}

%In Fig. \ref{evsTm.eps} the various transformation temperatures depicted in Table \ref{results} are plotted as a function of e/a ratio. It is interesting to note that this plot is quite similar to the one observed in case of Ni$_2$Mn$_{1+x}$In$_{1-x}$ indicating similarity of driving forces responsible for these transformations in the two series of alloys \cite{Planes,Sutou}.

%\begin{figure}
%\centering
%\includegraphics[width=\columnwidth]{evsTm.eps}
%\caption{\label{evsTm.eps} relationship between T$_C$,T$_M$ with e/a}
%\end{figure}

The austenitic state in these alloys is characterized by ferrimagnetic ordering due to antiparallel alignment of the two unequal Mn moments \cite{Chakrabarti,Luo}. However, the nature magnetic ordering in martensitic phase is not clear. It has been argued that martensitic transformation in such Mn rich alloys leads to change from a ferrimagnetic to an antiferromagnetic state with competing magnetic interactions \cite{nayak}. In the alloy compositions studied here, a near destruction of magnetic order is observed upon martensitic transformation followed by some kind of magnetic reordering at T$_C^M$. In order to understand in detail, the nature of magnetic interactions in the martensitic phase, isothermal magnetization (M(H)) measurements have been performed at several temperatures below T$_M$ of Mn$_2$Ni$_{1.6}$In$_{0.4}$ and Mn$_{1.96}$Ni$_{1.64}$In$_{0.4}$ and compared them with structurally stable Mn$_{2.04}$Ni$_{1.56}$In$_{0.4}$ alloy.

\begin{figure}
\centering
\includegraphics[width=75mm]{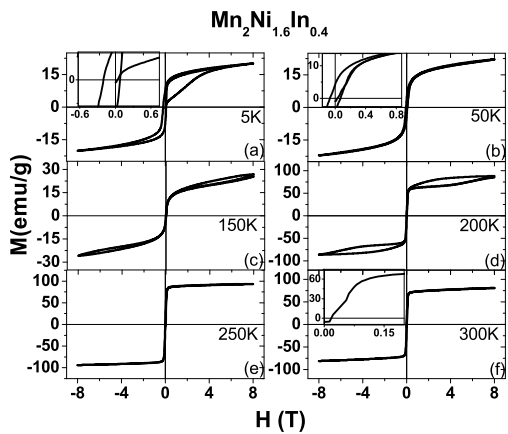}
\caption{\label{MH.eps} MH plot for  Mn$_{2}$Ni$_{1.6}$In$_{0.4}$ at (a) 5K, (b) 50K, (c) 150K, (d) 200K, (e) 250K and (f) 300K}
\end{figure}

The M(H) data for Mn$_{2}$Ni$_{1.6}$In$_{0.4}$ at various temperatures from 5K to 300K is presented in Fig. \ref{MH.eps}. The M(H) curves at 300K and 250K are in the ferrimagnetic austenitic state. A metamagnetic transition at very low values of magnetic field and the non-saturating magnetic moments are characteristics of these hysteresis loops (See inset of Fig. \ref{MH.eps}(f)) . These characteristics signify the presence of a weak antiferromagnetic interactions \cite{Ali}. It may be noted that Ni too has a magnetic moment, though much smaller than Mn, can interact with the ordered moment of Mn in presence of magnetic field giving rise to a metamagnetic transition. Furthermore, there are preferential site occupancies of various atoms forming Heusler compositions \cite{GDLiu}. Accordingly, in Mn$_2$Ni$_{1.6}$In$_{0.4}$, it is possible that Ni may prefer the X sites forcing Mn to occupy the In (Z) sites. This will lead to new magnetic correlations resulting in complex magnetic order.  

In the martensitic state the competing interactions are further amplified as can be seen from the M(H) loop at 200K (Fig. \ref{MH.eps}). The hysteresis loop is similar to that at 250K curve but with lobular structures in the positive and negative applied field regions extending over the entire range of magnetic field. This is due to non reversibility of magnetic moments arising from formation of new martensitic variants in presence of magnetic field. These lobular structure in the M(H) loops disappear below 150K or for the loops recorded below T$_C^M$. The M(H) loops recorded at T $<$ 100K exhibit finite coercivity indicating a build up of ferromagnetic interactions. Such ferromagnetic nanodomains have also been seen in Mn$_2$NiGa and are ascribed to antisite disorder \cite{singh}.  The M(H) loop at 50K has a coercivity of around 695 Oe and it increases with decreasing temperature. The hysteresis loops, though recorded under zero field cooling protocol, exhibit exchange bias effect which have been recently demonstrated to be characteristic of ferrimagnetic order. This effect can be readily seen at 5K and the calculated exchange bias field H$_{EB}$ is -639 Oe while the coercivity H$_{c}$ is $\sim$ 1400 Oe. It can also be seen that the virgin magnetization curve lies outside the hysteresis envelope indicating presence of more than one magnetic phases below T$_C^M$.

\begin{figure}
\centering
\includegraphics[width=\columnwidth]{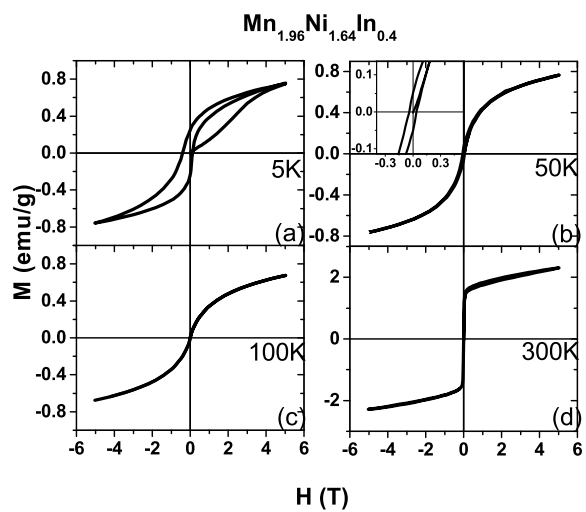}
\caption{\label{Mn196MH.eps} MH data for  Mn$_{1.96}$Ni$_{1.64}$In$_{0.4}$ Figs(a)to(d) are the plots for
temperatures (a) 5K,(b) 50K,(c) 100K and (d) 300K respectively}
\end{figure}

Nearly similar behavior of M(H) loops is noticed in case of Mn$_{1.96}$Ni$_{1.64}$In$_{0.4}$ as can be seen from Fig. \ref{Mn196MH.eps}. Here too, the zero field cooled hysteresis loops exhibit exchange bias effect. The estimated values of coercive field and exchange bias field are  H$_c$ = 2784 Oe and H$_{EB}$= 1329 Oe  respectively. These values are much higher than those calculated for Mn$_2$Ni$_{1.6}$In$_{0.4}$. This is again an interesting observation because it implies that with increase in Ni concentration both, ferromagnetic and antiferromagnetic interaction strengthen. This is possible only if increase in Ni concentration is in some way responsible for more antisite disorder.

\begin{figure}
\centering
\includegraphics[width=\columnwidth]{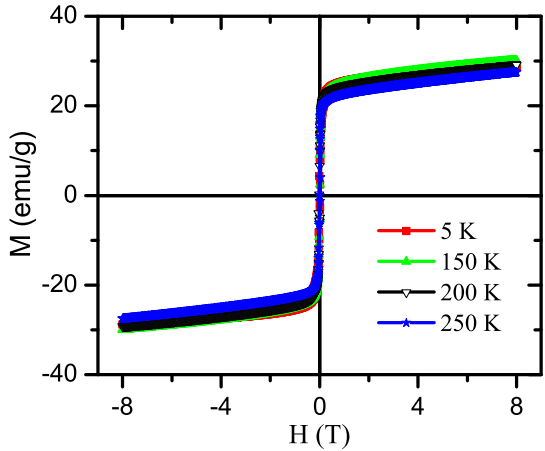}
\caption{\label{Mn204MH.eps} MH data for  Mn$_{2.04}$Ni$_{1.56}$In$_{0.4}$ }
\end{figure}

In Fig. \ref{Mn204MH.eps}, M(H) loops for Mn$_{2.04}$Ni$_{1.56}$In$_{0.4}$ are presented. It is interesting to see that a slight decrease in Ni concentration not only results a stable austenitic structure. All the M(H) loops recorded from 5K to 300K exhibit similar behavior with no saturation of magnetic moment up to H = 7T. However, the magnetic moment value recorded is quite low $\sim$ 32 emu/gm and is nearly independent of temperature.

\section{Discussion}
Mn$_2$NiIn is a Heusler alloy wherein the X sites are shared equally by Ni and Mn, the Y sites are occupied entirely by Mn and the Z sites by In. Increasing Ni content at the expense of In concentration leads to
martensitic alloys. The martensitic transformation occurs at nearly the same value of e/a as in case of Mn rich Ni$_2$MnIn type alloys. Another interesting aspect of this study is the critical dependence of martensitic and magnetic properties on Ni concentration. Though all the alloys exhibit ferrimagnetic ordering in the austenitic state, the martensitic state is characterized with the  presence of ferromagnetic and antiferromagnetic interactions. Both these interactions seem to dominate with increase in Ni concentration.

From the individual site preferences of transition metal ions, it is expected that Ni will prefer X sites forcing Mn in occupying the Z sites. Therefore an increase in Ni concentration will lead to more Mn occupying Z sites resulting in an antisite disorder. Such an antisite disorder is known to induce competing magnetic interactions result in a complex magnetic ground state \cite{singh, nayak}. Indeed, the martensitic state of these alloys exhibit strongly competing magnetic interactions. In order to determine if site preferences of various atoms is responsible for the observed magnetic ground state, we compare the values of magnetic moment at 300K in Mn$_2$Ni$_{1.6}$In$_{0.4}$ and its Ni rich ($y$ = 0.04 and 0.08) and Ni deficient ($y$ = -0.04 and -0.08) counterparts. Though both Ni deficient
and Ni deficient alloys have similar ferrimagnetic ordering temperature, a comparison of M(H) loops at 300K shows that magnetic moment measured at 7T in Mn$_{2.04}$Ni$_{1.56}$In$_{0.4}$ ($\sim$32 emu/g) is less than half of that in Mn$_2$Ni$_{1.6}$In$_{0.4}$ ($\sim$80 emu/g). Such a large decrease in the value of magnetic moment can only be accounted for by considering a strong antiferromagnetic component coming from a sizable fraction of Mn occupying Z sites. The sizable shift of Mn atoms at Z sites reduces its occupancy of X sites. Such a site occupancy disorder results in increasing the dominance of both, ferromagnetic interaction between Mn atoms in Y sublattice and antiferromagnetic interactions between Mn(Y) and Mn(Z) atoms. Such a competition then gives rise to the  observed strong exchange bias effect.

\section{Conclusion}
In summary, we have presented a systematic study of the martensitic and magnetic interactions in Heusler alloys of the type Mn$_2$Ni$_{(1+x)}$In$_{(1-x)}$ and Mn$_{(2-y)}$Ni$_{(1.6+y)}$In$_{0.4}$.  Martensitic instability can be induced by increasing Ni content at the expense of In. Such a increase in Ni content results in a site occupancy disorder due to preferential occupation of X sites by Ni atoms forcing Mn atoms to occupy the Z sites. This results in a transformation of magnetic ground state from ferrimagnetic to one dominated by presence of ferromagnetic Mn(Y)-Mn(Y) and antiferromagnetic Mn(Y)-Mn(Z) interactions. Such a transition is responsible for observation of properties like exchange bias effect in zero field cooled state in these Mn rich martensitic alloys.

\section*{Acknowledgements}

Authors would like to acknowledge the financial assistance from Council for Scientific and Industrial Research, New Delhi under the project 09/EMR-II/1188. Help of Devendra Buddhikot and Ganesh Jangam in magnetization measurements is gratefully acknowledged.

\end{document}